\documentstyle[12pt]{article}
\begin{document}
\newcommand {\ee}{\end{equation}}
\newcommand {\bea}{\begin{eqnarray}}
\newcommand {\eea}{\end{eqnarray}}
\newcommand {\nn}{\nonumber \\}
\newcommand {\Tr}{{\rm Tr\,}}
\newcommand {\tr}{{\rm tr\,}}
\newcommand {\e}{{\rm e}}
\newcommand {\etal}{{\it et al.}}
\newcommand {\m}{\mu}
\newcommand {\n}{\nu}
\newcommand {\pl}{\partial}
\newcommand {\p} {\phi}
\newcommand {\vp}{\varphi}
\newcommand {\vpc}{\varphi_c}
\newcommand {\al}{\alpha}
\newcommand {\be}{\beta}
\newcommand {\ga}{\gamma}
\newcommand {\Ga}{\Gamma}
\newcommand {\x}{\xi}
\newcommand {\ka}{\kappa}
\newcommand {\la}{\lambda}
\newcommand {\La}{\Lambda}
\newcommand {\si}{\sigma}
\newcommand {\th}{\theta}
\newcommand {\Th}{\Theta}
\newcommand {\om}{\omega}
\newcommand {\Om}{\Omega}
\newcommand {\ep}{\epsilon}
\newcommand {\vep}{\varepsilon}
\newcommand {\na}{\nabla}
\newcommand {\del}  {\delta}
\newcommand {\Del}  {\Delta}
\newcommand {\mn}{{\mu\nu}}
\newcommand {\ls}   {{\lambda\sigma}}
\newcommand {\ab}   {{\alpha\beta}}
\newcommand {\half}{ {\frac{1}{2}} }
\newcommand {\fourth} {\frac{1}{4} }
\newcommand {\sixth} {\frac{1}{6} }
\newcommand {\sqg} {\sqrt{g}}
\newcommand {\fg}  {\sqrt[4]{g}}
\newcommand {\invfg}  {\frac{1}{\sqrt[4]{g}}}
\newcommand {\sqZ} {\sqrt{Z}}
\newcommand {\gbar}{\bar{g}}
\newcommand {\sqk} {\sqrt{\kappa}}
\newcommand {\sqt} {\sqrt{t}}
\newcommand {\reg} {\frac{1}{\epsilon}}
\newcommand {\fpisq} {(4\pi)^2}
\newcommand {\Lcal}{{\cal L}}
\newcommand {\Ocal}{{\cal O}}
\newcommand {\Dcal}{{\cal D}}
\newcommand {\Ncal}{{\cal N}}
\newcommand {\Mcal}{{\cal M}}
\newcommand {\scal}{{\cal s}}
\newcommand {\Dvec}{{\vec D}}
\newcommand {\dvec}{{\vec d}}
\newcommand {\Evec}{{\vec E}}
\newcommand {\Hvec}{{\vec H}}
\newcommand {\Vvec}{{\vec V}}
\newcommand {\Btil}{{\tilde B}}
\newcommand {\ctil}{{\tilde c}}
\newcommand {\Stil}{{\tilde S}}
\newcommand {\Ztil}{{\tilde Z}}
\newcommand {\altil}{{\tilde \alpha}}
\newcommand {\betil}{{\tilde \beta}}
\newcommand {\latil}{{\tilde \lambda}}
\newcommand {\phitil}{{\tilde \phi}}
\newcommand {\Phitil}{{\tilde \Phi}}
\newcommand {\natil} {{\tilde \nabla}}
\newcommand {\ttil} {{\tilde t}}
\newcommand {\Shat}{{\hat S}}
\newcommand {\shat}{{\hat s}}
\newcommand {\xhat}{{\hat x}}
\newcommand {\Zhat}{{\hat Z}}
\newcommand {\Gahat}{{\hat \Gamma}}
\newcommand {\nah} {{\hat \nabla}}
\newcommand {\gh}  {{\hat g}}
\newcommand {\labar}{{\bar \lambda}}
\newcommand {\cbar}{{\bar c}}
\newcommand {\bbar}{{\bar b}}
\newcommand {\Bbar}{{\bar B}}
\newcommand {\psibar}{{\bar \psi}}
\newcommand {\chibar}{{\bar \chi}}
\newcommand {\bbartil}{{\tilde {\bar b}}}
\newcommand {\intfx} {{\int d^4x}}
\newcommand {\change} {\leftrightarrow}
\newcommand {\ra} {\rightarrow}
\newcommand {\larrow} {\leftarrow}
\newcommand {\ul}   {\underline}
\newcommand {\pr}   {{\quad .}}
\newcommand {\com}  {{\quad ,}}
\newcommand {\q}    {\quad}
\newcommand {\qq}   {\quad\quad}
\newcommand {\qqq}   {\quad\quad\quad}
\newcommand {\qqqq}   {\quad\quad\quad\quad}
\newcommand {\qqqqq}   {\quad\quad\quad\quad\quad}
\newcommand {\qqqqqq}   {\quad\quad\quad\quad\quad\quad}
\newcommand {\qqqqqqq}   {\quad\quad\quad\quad\quad\quad\quad}
\newcommand {\lb}    {\linebreak}
\newcommand {\nl}    {\newline}

\newcommand {\vs}[1]  { \vspace*{#1 cm} }

\newcommand {\MPL}  {Mod.Phys.Lett.}
\newcommand {\NP}   {Nucl.Phys.}
\newcommand {\PL}   {Phys.Lett.}
\newcommand {\PR}   {Phys.Rev.}
\newcommand {\PRL}   {Phys.Rev.Lett.}
\newcommand {\CMP}  {Commun.Math.Phys.}
\newcommand {\AP}   {Ann.of Phys.}
\newcommand {\PTP}  {Prog.Theor.Phys.}
\newcommand {\NC}   {Nuovo Cim.}
\newcommand {\CQG}  {Class.Quantum.Grav.}


\font\smallr=cmr5
\def\ocirc#1{#1^{^{{\hbox{\smallr\llap{o}}}}}}
\def\ogamma{\ocirc{\gamma}{}}
\def\oM{{\buildrel {\hbox{\smallr{o}}} \over M}}
\def\osigma{\ocirc{\sigma}{}}

\def\overleftrightarrow#1{\vbox{\ialign{##\crcr
 $\leftrightarrow$\crcr\noalign{\kern-1pt\nointerlineskip}
 $\hfil\displaystyle{#1}\hfil$\crcr}}}
\def\overnab{{\overleftrightarrow\nabslash}}

\def\va{{a}}
\def\vb{{b}}
\def\vc{{c}}
\def\tilpsi{{\tilde\psi}}
\def\tbpsi{{\tilde{\bar\psi}}}

\def\Dslash{{}\hbox{\hskip2pt\vtop
 {\baselineskip23pt\hbox{}\vskip-24pt\hbox{/}}
 \hskip-11.5pt $D$}}
\def\nabslash{{}\hbox{\hskip2pt\vtop
 {\baselineskip23pt\hbox{}\vskip-24pt\hbox{/}}
 \hskip-11.5pt $\nabla$}}
\def\xislash{{}\hbox{\hskip2pt\vtop
 {\baselineskip23pt\hbox{}\vskip-24pt\hbox{/}}
 \hskip-11.5pt $\xi$}}
\def\leftnabla{{\overleftarrow\nabla}}

\def\delL{{\delta_{LL}}}
\def\delG{{\delta_{G}}}
\def\delc{{\delta_{cov}}}


\begin{flushright}
US-97-08
\end{flushright}

\vs1

\begin{center}
{\Large\bf General Structure of Conformal Anomaly 
and 4 Dimensional Photon-Dilaton Gravity }

\vspace{2cm}

{\large Shoichi ICHINOSE
\footnote{
e-mail : ichinose@u-shizuoka-ken.ac.jp}
}
\vspace{1cm}

{\large Department of Physics \\
Universuty of Shizuoka \\
Yada 52-1, Shizuoka 422-8526, JAPAN\\}

\end{center}
\vfill
{\large Abstract}\nl 
The general structure of the conformal anomaly
and the dilaton's effect to the anomaly
are analysed. First we give a new formal 
proof of the statement that the conformal anomaly,
in the theory which is conformal
invariant at the classical level, 
is conformal invariant. The heat-kernel
regularization and Fujikawa's method are taken for
the analysis. We
present a new explicit result of the conformal anomaly in
4 dimensional photon-dilaton gravity. This
result is examined from the point of the general structure.

\newpage
\section{Introduction}
The conformal anomaly is a common phenomenon
in the quantum field theories. It means the conformal
symmetry is inevitably broken by the quantization
procedure even if we start from the conformally
invariant theory at the classical level. 
The quantization generally
requires some mass scale in order to regularize
the theory, which causes the anomaly. Many explicit
calculations have been performed and they
indicate the existence of some structure in
the anomaly. 
It has been noticed and analysed since a long time ago\cite{MD77}.
Several years ago, Deser and Schwimmer\cite{DS93}
argued that the conformal anomaly are made of three
types of terms:\ A)\ the Euler term;\ B)\ the conformal
invariants;\ C)\ "trivial" terms. They took the dimensional
regularization and analysed the effective action at the 
1-loop perturbative calculation.
The formal proof, however, does not seem to exist so far. 
We examine the problem in a different approach. Firstly we adopt
the heat-kernel regularization
for the ultraviolet divergences. Secondly the anomaly is taken from
the Jacobian factor under the change of integration variables
(Fujikawa's method\cite{KF79,KF85}). It is known that this method makes
the anomaly calculation clear and all known anomalies are
succinctly reproduced\cite{II96,DAMTP9687}. 
Based on these, we give a formal proof about the
conformal invariance of the conformal anomaly up to 
"trivial" terms(defined in Sec.2).

We present an explicit new result of the conformal
anomaly in the photon-dilaton gravity. The
dilaton and the gravitational fields are treated as the background
field. We examine the result from the point of the conformal symmetry.
Although the dilaton is a popular field in the string theory,
its true role in the field theory still remains obscure. 
It has the characteristic interaction of the type 
$\e^{\mbox{const}\times\p}$. 
Recently, in the 2 dimensional(dim) simple model, some interesting aspect
of the dilaton is clarified in relation to 
the Weyl anomaly and the Hawking radiation
\cite{BH,NO6143,SI97}.
In the present paper, we examine the 4 dim case. 
It turns out that the dilaton field behaves in good harmony with
the gravitational field.

\section{Conformal Invariance of the Conformal \\
Anomaly}

Let us consider a conformally invariant action $S[\vp;g]$ 
in the $n$-dim Euclidean space with the following quadratic structure
with respect to (w.r.t.) a matter field.
\bea
\label{CI.1}
\e^{\Ga[g]}=\int\Dcal\vp~\e^{S[\vp;g]}\com\q
S[\vp;g]=\int d^nx~\vp\Dvec_g\vp\com
\eea
where $\vp$ and $g_\mn$ are the quantum matter field
(appropriately scaled, we consider the scalar field for
simplicity) and the background metric respectively. 
$\Ga[g]$ is the Wilsonian effective action.
$\Dvec_g$ is
an elliptic differetial operator of the system (we call
it the {\it system operator}). Free fields (scalar,spinor,vector, etc.), 
on the curved background space,  
all have the above structure essentially. 
$S[\vp;g]$
is assumed to be invariant under the Weyl (conformal) 
transformation.
\bea
\label{CI.2}
\vp'=W_\al(x)\vp\com\q {g_\mn}'=\e^{2\al(x)}g_\mn\com\nn
S[\vp';g']=S[\vp;g]\com\q W_\al\Dvec_{g'}W_\al=\Dvec_g\com
\eea
where $\al(x)$ is the Weyl transformation parameter. Generally
$W_\al$ has the form $W_\al(x)\propto \e^{p\al(x)}$, with a constant
$p$\ depending on each matter field.
\footnote{
The suffix '$\al$' of $W_\al$ shows its dependence on $\al(x)$, not
the field index.
} 

Let us evaluate the variation
of $\Ga[g]$, under the Weyl transformation, as
\bea
\label{CI.3}
\e^{\Ga[g']+\Del S[g']}
=\int\Dcal\vp'\e^{S[\vp';g']+\Del S[g']}
=\int\Dcal\vp \det\frac{\pl\vp'}{\pl\vp}\e^{S[\vp;g]+\Del S[g']}\com
\eea
where we have introduced the counterterm action $\Del S[g]$ which
depends only on the background metric.
Following Fujikawa, we identify the Jacobian\ 
$\det\frac{\pl\vp'}{\pl\vp}=\exp\Tr \ln\{W_\al(x)\del^n(x-y)\}
=\exp\Tr\{p~\al(x)\del^n(x-y)+O(\al^2)\}
$\ 
as the conformal anomaly and regularize it by the heat-kernel
for the system operator $\Dvec_g$.
\bea
\label{CI.4}
\Ga[g']+\Del S[g']-\Ga[g]-\Del S[g]=p~\Tr \al(x)\del^n(x-y)
+\Del S[g']-\Del S[g]
+O(\al^2)\nn
= \lim_{t\ra +0}\{ p~\Tr\al(x)<x|\e^{-t\Dvec_g}|y>
+\Del S[g']-\Del S[g]\}
+O(\al^2)\com
\eea
where $\al(x)$ is considered infinitesimal and $t$ is the proper
time which plays the role of the ultraviolet regularization.
Finally the conformal anomaly is given by
\bea
\label{CI.5}
\mbox{Anomaly}\equiv
\frac{\del}{\del\al(x)}(\Ga[g']+\Del S[g']-\Ga[g]-\Del S[g])|_{\al=0}\nn
= \lim_{t\ra +0}\{ p~\tr G_g(x,x;t)+
\frac{\del}{\del\al(x)}\Del S[g']|_{\al=0}\}\ ,\ 
G_g(x,y;t)=<x|\e^{-t\Dvec_g}|y>\ ,\nn
(\frac{\pl}{\pl t}+\Dvec_g(x))G_g(x,y;t)=0\com\q
\lim_{t\ra +0}G_g(x,y;t)=\del^n(x-y)\com
\eea
where  $\tr G_g(x,x;t)$ has pole terms proportional to
$1/t^s$\ ($s=n/2,n/2-1,\cdots,1$) as $t\ra +0$, 
which are subtracted by
$\Del S$-terms through 
$\frac{\del}{\del\al(x)}\Del S[g']|_{\al=0}$, i.e.,
the renormalization of some massive parameters such as
the cosmological constant and the gravitational constant.
The $t^0$-part of $\tr G_g(x,x;t)$ is the conformal anomaly.
Some terms of it can be subtracted by the finite part of
$\Del S$ through 
$\frac{\del}{\del\al(x)}\Del S[g']|_{\al=0}$.
They are called "trivial" (anomaly) terms.
("Trivial" terms are defined by those which can be
obtained by the Weyl transformation of local actions.)
It is the advantage of the present approach that
the anomaly is so succinctly expressed as in (\ref{CI.5}). We will now
prove the conformal invariance of the anomaly using this
expression.

In the heat-kernel equation of (\ref{CI.5}), 
first we replace $g_\mn$ with the Weyl-
transformed one ${g_\mn}'$ of (\ref{CI.2})
and then we multiply it by $W_\al(x)$
from "left" and by $W_\al(y)$ from "right".
\bea
\label{CI.6}
(W_\al(x)^2\frac{\pl}{\pl t}+W_\al(x)\Dvec_{g'}(x)W_\al(x))
W_\al(x)^{-1}G_{g'}(x,y;t)W_\al(y)=0\com
\eea
where $W_\al(x)\Dvec_{g'}(x)W_\al(x)$ is the Weyl-transformed
operator, therefore the operand 
$W_\al(x)^{-1}G_{g'}(x,y;t)W_\al(y)\equiv G'(x,y;t)$
can be considered to be the Weyl-transformed heat-kernel. 
In particular the transformed heat-kernel $G'(x,y;t)$ satisfies
the boundary condition:\ 
$\lim_{t\ra +0}G'(x,y;t)=\del^n(x-y)$.
Using
the conformal invariance of the system operator
$W_\al(x)\Dvec_{g'}(x)W_\al(x)=\Dvec_{g}(x)$ of (\ref{CI.2}), $G'$
satisfies
\bea
\label{CI.7}
(W_\al(x)^2\frac{\pl}{\pl t}+\Dvec_g(x))G'(x,y;t)=0\pr
\eea
Introducing a new point $z$ near $x$, we can rewrite Eq.(\ref{CI.7})
as
\bea
\label{CI.7b}
z\ra x\q,\q
(W_\al(z)^2\frac{\pl}{\pl t}+\Dvec_g(x))G'(x,y;t)\nn
=(W_\al(z)^2-W_\al(x)^2)\frac{\pl}{\pl t}G'(x,y;t)
=f(z,x)\frac{\pl}{\pl t}G'(x,y;t)\com
\eea
where $f(z,x)\equiv W_\al(z)^2-W_\al(x)^2,\ f(x,x)=0.$
Instead of (\ref{CI.7}), we solve (\ref{CI.7b}) for $z\ra x$.
Because $\lim_{z\ra x}f(z,x)=0$, we can treat the term of
$f(z,x)\frac{\pl}{\pl t}G'(x,y;t)$ as a small perturbation
so far as $x\neq y$.
Comparing the heat-kernel equations in
(\ref{CI.5}) and (\ref{CI.7b}), we see
\bea
\label{CI.8}
G'(x,y;t)=\lim_{z\ra x} \{ G_g(x,y;\frac{t}{W_\al(z)^2})
+O(f) \}\com
\eea
where $O(f)$ means the contribution from
$f(z,x)\frac{\pl}{\pl t}G'(x,y;t)$.
If we take the limit $|z-x|\ra +0$ before the limit $|x-y|\ra +0$,
the $O(f)$ term vanish.
\footnote{
We can regard this limitting procedure as a sort of (ultraviolet)
regularization.
}
In this case, 
this equation shows that 
$t^0$-part of $\tr G_g(x,x;t)$, which is the conformal anomaly up to
"trivial" terms, is conformal-invariant.
\footnote{  
This proof is contrasting with the case of "gauge-invariance" of
the heat-kernel\cite{DAMTP9687} where the invariance is valid
for all powers of $t$.
}
It is well-established that $t^0$-part of $\tr G_g(x,x;t)$
corresponds to the log-divergent part of the quantum system of $\Dvec$
, say, in the dimensional regularization, so this result is
consistent with Ref.\cite{DS93}'s characterization of B-type terms.
So far we have analysed only local properties of $\tr G_g(x,x;t)$.
We can not exclude the topological term (Euler term) in it,
because it does not depend on any local quantity.
This completes a formal proof of the general structure of Weyl anomaly
stated in Sec.1.
The explicit calculations, such as the scalar-gravity theory, 
show its validity.

In the case of the gauge theory on the curved background,
the above proof can not be applied directly because
we have to introduce the gauge-fixing term and which
breaks the conformal invariance of the action.
\footnote{
From the point of conformal group symmetry on the
flat space ( SO(4,2)-symmetry ), it was already noted that the gauge-fixing
procedure requires some nonlocal terms in the
BRST transformation if the gauge
theory is BRST-quantized keeping the conformal symmetry
of the action at the classical level.
\cite{FP84, SI86LMP,SI86NP}
}
We notice, however, the gauge-fixed action combined with the
ghost term is conformal invariant up to a "BRST-trivial" term (see Sec.3).
Therefore we can still expect that the conformal invariance is kept
for the case of the gauge theory. 
Explicit calculations of some theories show
it is indeed the case. The conformal anomaly of the 4 dim free photon
on the curved background was obtained by \cite{BC77,DK77} 
in the zeta-function
regularization as
\bea
\label{CI.21}
\mbox{Anomaly(photon,no dilaton)}=\frac{1}{(4\pi)^2}
\frac{1}{180}(-31\sqg E+18I_0+18J_0)\pr
\eea
It is made of the Euler term $\sqg E$ with
\bea
\label{CI.22}
E=\fourth R_{\mn\ab}R_{\ls\ga\del}\ep^{\mn\ls}\ep^{\ab\ga\del}
=R^2+R_{\mn\ab}R^{\mn\ab}-4R_\mn R^\mn\com
\eea
and a conformal invariants $I_0$ defined by
\bea
\label{CI.23}
I_0\equiv\sqg C_{\mn\ls}C^{\mn\ls}
=\sqg (-2R_\mn R^\mn
+R_{\mn\ls}R^{\mn\ls}+\frac{1}{3}R^2),
I_0'=I_0 \com
\eea
where $C_{\mn\ls}$ is the Weyl tensor, 
and a "trivial" term $J_0$ defined by
\bea
\label{CI.24}
J_0\equiv\sqg\na^2R\com\q
K_0\equiv\sqg R^2\com\q
g_\mn\frac{\del}{\del g_\mn (x)}\intfx K_0=6 J_0
\pr
\eea

\section{Trace anomaly of 4D free photon on the background
dilaton-gravity}

It is very interesting to know how the dilaton affects the
structure of the conformal anomaly.
The non-gauge case
was already analysed in Ref.\cite{NO6143}
where the scalar-dilaton-gravity theory, 
$
S_S =\half\intfx\sqg \e^{P(\p)}\vp (-\na^2-\frac{1}{6} R)\vp
$,
($\p$:\ dilaton;\ $\vp$:\ scalar;\ $P(\p)$ is an arbitrary
function of $\p$) is taken.
Using 
another conformal invariant $I_4$ defined by
\bea
\label{s2.12b}
I_4\equiv\sqg(P_{,\m}P^{,\m})^2 \com\q
I_4'=I_4\com
\eea
and another "trivial" term $J_{2a}$ defined by
\bea
\label{s2.12c}
J_{2a}\equiv\sqg\na^2(P_{,\m}P^{,\m})\ ,\ 
K_{2a}\equiv\sqg R P_{,\m}P^{,\m}\ ,\ 
g_\mn\frac{\del}{\del g_\mn(x)}\intfx K_{2a}=+3 J_{2a}\ ,
\eea
besides $\sqg E, I_0, J_0$ defined in Sec.2, 
the anomaly is given by
\bea
\label{s2.12d}
\mbox{Anomaly(scalar)}=\frac{1}{(4\pi)^2}\{
\frac{1}{360}(3I_0-\sqg E)+\frac{1}{32}I_4
-\frac{1}{180}J_0-\frac{1}{24}J_{2a} \}\ .
\eea
The general structure given in Sec.2 is strictly obeyed
in this case. 
In the notation $I_0,I_4,\cdots$, in the above and in Sec.2, 
the numbers in the lower suffixes 
show the power numbers of $P$. 

Let us consider the photon coupled system, instead of the scalar field, 
and examine whether the anomaly keeps this general structure or not.
This is the gauge theory and we expect a new aspect will appear
due to the gauge-fixing procedure.
The action is given by
\bea
\label{s1}
&S_V=\int d^4x \sqrt{g}(-{1 \over 4}\e^{P(\phi)}F_\mn F^\mn )\ ,\ 
F_\mn=\na_\m A_\n-\na_\n A_\m&
\eea
Besides the 4 dim general coordinate symmetry, 
this theory has the local Weyl symmetry, 
\bea
\label{s1a}
g_\mn'=\e^{2\al(x)}g_\mn\com\q 
A_\m:\ \mbox{fixed}\com\q
\p:\ \mbox{fixed}\com
\eea
and has the Abelian local gauge symmetry.
\bea
\label{s1b}
A_\m'=A_\m+\na_\m\La(x)\com\q
g_\mn:\ \mbox{fixed}\com\q \p:\ \mbox{fixed}\pr
\eea

The gauge-fixing term and the corresponding ghost lagrangian
are
\bea
\label{s2}
S_{gf}=\int d^4x \sqrt{g}(-\half \e^{P(\phi)}(\na_\m A^\m)^2 )\ ,\ 
S_{gh}=\intfx\sqg \e^{P(\p)}\cbar\na_\m \na^\m c\ ,
\eea
where $\cbar$ and $c$ are the anti-ghost and ghost fields
respectively. They  are hermitian scalar fields with Fermi statistics
(Grassmann variables). The total action 
$S[A,\cbar,c;g,\p]=S_V+S_{gf}+S_{gh}$, is invariant for the
BRST symmetry.
\bea
\label{s2b}
\del A_\m=\xi\pl_\m c\equiv\xi~\shat A_\m\com\q 
\del\cbar=\xi\na^\m A_\m\equiv\xi~\shat\cbar\com\nn
\del c=0\equiv\xi~\shat c\com\q 
\del\p=0\equiv\xi~\shat\p\com\q 
\del g_\mn=0\equiv\xi~\shat g_\mn\com
\eea
where $\xi$ is the BRST parameter (Grassmann,global)
and $\shat$ is introduced as the BRST operator. 
$S_{gf}$\ breaks the Weyl symmetry (\ref{s1a}), besides 
the gauge symmetry (\ref{s1b}).
Therefore we must take close care to define the conformal anomaly
generally in the gauge theories. It is non-trivial
whether the general structure suggested by Ref.\cite{DS93}
holds true.

We can fix the Weyl transformation of $\cbar$ and $c$ 
\cite{KF81} as
\bea
\label{s2c}
\cbar'=\e^{-2\al(x)}\cbar\com\q c'=c\com
\eea
from two requirements:\ 
(i)\ the ghost action $S_{gh}$, (\ref{s2}), 
is invariant for the global Weyl transformation ($\al(x)$ is
independent of $x$);\ 
(ii)\ For the {\it infinitesimal} Weyl transformation, 
$S_{gf}+S_{gh}$ is conformal invariant up to "BRST-trivial" terms.
\bea
\label{s2d}
(S_{gf}+S_{gh})'-(S_{gf}+S_{gh})=\intfx\sqg\e^{P(\p)}
\{-2~\shat (\cbar\na^\la\al\cdot A_\la)+O(\al^2)\}\ .
\eea

After rescaling the vector and ghost fields 
( for normalizing the kinetic term):\ 
$ \e^{\half P(\p)}A_\m=B_\m\ ,\ \e^{P(\p)}\cbar=\bbar$,
the actions can be expressed as
\bea
\label{s3}
&S_V+S_{gf}=\int d^4x [
\half \sqg B^\m \{g_\mn \na^2+(\Ncal^\la)_\mn\na_\la+\Mcal_\mn\} B^\n  
+\mbox{total deri.}]\ ,&\nn
&(\Ncal^\la)_\mn=P_{,\m}\del^\la_\n-P_{,\n}\del^\la_\m\com\q
\Mcal_\mn=R_\mn-\fourth g_\mn (P_{,\la}P^{,\la}+2\na^2P)\com &\nn
&S_{gh}=\intfx\sqg\bbar\na^2c\pr &
\eea
As shown by Fujikawa\cite{KF83}, we should adopt 
the BRST(w.r.t. 4 dim general
coordinate transformation)-invariant measure by taking
$\Btil_\m,\bbartil$ and $\ctil$, instead of $B_\m, \bbar$ and $c$
respectively, as the path-integral variables.
\footnote{
The measure is also invariant under the BRST transformation
w.r.t. the Abelian gauge symmetry:
$
\del\Btil_\m=\xi\e^{P/2}g^{-1/8}(\pl_\m\ctil
           -\fourth g^{-1}\pl_\m g\cdot\ctil)\ ,\ 
\del\bbartil=\xi\e^{P/2}g^{1/8}(\na^\m\Btil_\m
-\frac{1}{8}g^{-1}g^{,\m}\Btil_\m-\half P^{,\m}\Btil_\m)\ ,\ 
\del\ctil=0\ ,\ 
\det\{
   \pl({\Btil_\m}',\bbartil',\ctil')/ \pl(\Btil_\n,\bbartil,\ctil)
    \}
=1.	   
$
}
\bea
\label{s3a}
\Btil_\m\equiv g^{1/8} B_\m\ \ (\Btil^\m\equiv g^{3/8} B^\m)\com
\bbartil\equiv\fg\bbar\com\ctil\equiv\fg c\com\nn
S_V+S_{gf}=\int d^4x [
-\half \Btil^\m\Dvec_\m^{~\n} \Btil_\n  +\mbox{total deri.}]\ ,\nn
\Dvec_\m^{~\n}
=-g^{1/8}\{\del_\m^{~\n} \na^2+(\Ncal^\la)_\m^{~\n}\na_\la
+\Mcal_\m^{~\n}\}g^{-1/8}\ ,\nn
S_{gh}=-\intfx \bbartil\ \dvec\ \ctil\com\q 
\dvec\equiv -\fg\na^2\invfg\com\nn
\exp\{\Ga[g,\p]\}=\int\Dcal\Btil\Dcal\bbartil\Dcal\ctil~
\exp (\Stil[\Btil,\bbartil,\ctil;g,\p])\com\nn
\Stil[\Btil,\bbartil,\ctil;g,\p]=S[A,\cbar,c;g,\p]\com
\eea
where $\Ga[g,\p]$ is the (Wilsonian) effective action.

The Weyl transformation of (\ref{s1a}) and that of 
the integration variables
($\Btil_\m'=\e^{\al}\Btil_\m,\ \bbartil'=\bbartil,
\ \ctil'=\e^{2\al}\ctil$) give us the Ward-Takahashi identity
for the Weyl transformation.
\bea
\label{s3b}
\exp\Ga[g',\p]=\int\Dcal\Btil'\Dcal\bbartil\Dcal\ctil'
\exp \Stil[\Btil',\bbartil,\ctil';g',\p]       \nn
=\int\Dcal\Btil\cdot\det\frac{\pl \Btil'}{\pl \Btil}\cdot
\Dcal\bbartil\Dcal\ctil\cdot\det\frac{\pl \ctil'}{\pl \ctil}
\exp \Stil[\Btil',\bbartil,\ctil';g',\p]       \nn
=\int\Dcal\Btil\Dcal\bbartil\Dcal\ctil\cdot
\det(\e^{\al}\del_\m^{~\n}\del^4(x-y))\cdot
\det^{-1}(\e^{2\al}\del^4(x-y))         \nn
\times\exp \left[
\Stil[\Btil,\bbartil,\ctil;g,\p]
+\del\Btil_\m\frac{\del\Stil}{\del\Btil_\m}
+\del\ctil\frac{\del\Stil}{\del\ctil}
+\del g_\mn\frac{\del\Stil}{\del g_\mn}+O(\al^2)
          \right]       \pr
\eea
Considering the infinitesimal transformation
($\del g_\mn=2\al g_\mn,\ \del\Btil_\m=\al\Btil_\m,\ 
\del\ctil=2\al\ctil$) in the above, and regularizing
the space delta-functions $\del^4(x-y)$ in terms of the heat-kernels,
we obtain
\bea
\label{s3c}
\Ga[g',\p]-\Ga[g,\p]=2\al g_\mn\frac{\del\Ga}{\del g_\mn}+O(\al^2)\nn
=\al\left<\Btil_\m\frac{\del\Stil}{\del\Btil_\m}\right>
+2\al\left<\ctil\frac{\del\Stil}{\del\ctil}\right>
+2\al\left<g_\mn\frac{\del\Stil}{\del g_\mn}\right>\nn
+\Tr \ln\{\e^{\al}<x|\e^{-t\Dvec_\m^{~\n}}|y>\}
-\Tr \ln\{\e^{2\al}<x|\e^{-t\dvec}|y>\}+O(\al^2)\com
\eea
where $t$ is the regularization parameter:\ $t\ra +0$.
The variation part of $\Stil$ above gives the "naive"
Ward-Takahashi identity, while the remaining two terms (terms with "Tr")
give the deviation from it. Therefore 
the last two terms can be regarded as the Weyl anomaly in this theory.
\bea
\label{s3d}
T_1\equiv\left.\frac{\pl}{\pl\al}
\Tr \ln\{\e^{\al}<x|\e^{-t\Dvec_\m^{~\n}}|y>\}\right|_{\al=0,t=0}\com\nn
T_2\equiv -\left.\frac{\pl}{\pl\al}
\Tr \ln\{\e^{2\al}<x|\e^{-t\dvec}|y>\}\right|_{\al=0,t=0}\pr
\eea

The anomaly formula for the differential operator 
$
\Dvec_\m^{~\n}
=-g^{1/8}\{\del_\m^{~\n} \na^2+(\Ncal^\la)_\m^{~\n}\na_\la
+\Mcal_\m^{~\n}\}g^{-1/8}
$\ 
with arbitrary general covariants of $\Ncal^\la$\ and $\Mcal$ is
, taking the heat-kernel regularization
and the Fujikawa method,
 given as
\bea
\label{s4}
T=\frac{1}{(4\pi)^2}\sqrt{g}
                \left\{\  
\mbox{Tr}[\frac{1}{6}D^2X+\half X^2
+\frac{1}{12}Y_\ab Y^\ab] \right.          \nn
\left. +4\times \frac{1}{180}
 (R_{\mn\ls}R^{\mn\ls}-R_\mn R^\mn +0\times R^2-\na^2R ) 
                \right\} \ ,\nn
X_\m^{~\n}=\Mcal_\m^{~\n}-\half (\na_\al \Ncal^\al)_\m^{~\n}
-\fourth (\Ncal^\al \Ncal_\al)_\m^{~\n}-\frac{1}{6}\del_\m^{~\n}R\ ,\nn
(Y_\ab)_\m^{~\n}=\left\{ \half (\na_\al\Ncal_\beta)_\m^{~\n}
+\fourth (\Ncal_\al \Ncal_\beta)_\m ^{~\n}-\al\change\beta\right\}
+R_{\ab\m}^{~~~~\n}\ ,\nn
(D_\al X)_\m^{~\n}=\na_\al X_\m^{~\n}+\half[\Ncal_\al,X]_\m^{~\n}\ ,\nn
(D^2X)_\m^{~\n}=\na^2X_\m^{~\n}+\half [\Ncal^\al,D_\al X]_\m^{~\n}
\eea
This formula 
is the same as the counter-term formula of
\cite{BV85,BOS92} which was
derived in the dimensional regularization.
Putting all explicit equations into the formula, 
the vector contribution to the Weyl anomaly, $T_1$, is finally given by
\bea
\label{s5}
T_1=\frac{1}{(4\pi)^2} \sqg\left[ 
\frac{1}{180}
 (-11 R_{\mn\ls}R^{\mn\ls}+86 R_\mn R^\mn -20 R^2 +6\na^2R) 
\right. \nn
+\{ \frac{1}{16}(P_{,\m}P^{,\m})^2+\frac{1}{12}\na^2(P_{,\m}P^{,\m})
    +\frac{5}{12}(\na^2P)^2-\frac{1}{6}P_{,\mn}P^{,\mn}   \nn
    +\frac{2}{3}R^\mn P_{,\m}P_{,\n}-\frac{1}{6}RP_{,\m}P^{,\m} 
\}                                           \nn
\left.
+\{ -\frac{1}{3}\na^4P-\frac{1}{6}(P_{,\mn}+\half g_\mn \na^2P)\cdot
     P^{,\m}P^{,\n}                          
	 +\frac{1}{3}(R^\mn-\half g^\mn R)P_{,\mn}
 \}
                      \right]\pr
\eea
Terms in the first big bracket (\{\ \}) show the even power part
of $P(\p)$, while those in the second show the odd power part.
The ghost contribution to the Weyl anomaly is obtained, 
by evaluating the heat-kernel with $\dvec\equiv -\fg\na^2\invfg$,
as
\bea
\label{s6}
T_2=\frac{1}{(4\pi)^2} \sqg
\left[ 
-2\times \frac{1}{180}
 (-6\na^2R+R_{\mn\ls}R^{\mn\ls}-R_\mn R^\mn +\frac{5}{2}R^2 )
                      \right]\pr
\eea
There are no P-terms in the ghost part. $T_2$ is the same as
eq.(B.11) of \cite{RE84}. 
$T_1+T_2$ gives the final result of the Weyl anomaly. It is given
by the Euler term($\sqg E$), five Weyl invariants
($I_4,I_2,I_{1a},I_{1b},I_0$)
and four trivial terms($J_3,J_{2a},J_{2b},J_0$) which are
defined and explained in the following.
\bea
\label{s7}
\mbox{Anomaly(photon)}=\frac{1}{(4\pi)^2}
\{
\frac{1}{180}(-31\sqg E+18I_0+18J_0) \nn
+\frac{1}{16}I_4+I_2+\frac{1}{3}(I_{1a}-I_{1b})
-\frac{1}{12}J_3+\frac{1}{12}J_{2a}+\frac{1}{3} J_{2b}
\}\com
\eea
The Weyl anomaly for the ordinary (without the dilaton) 
photon-graviton theory (Sec.2) is correctly given
by the case of $P(\p)=0$.

In the above, we have new conformal invariants
$I_2,I_{1a},I_{1b}$ defined by
\bea
\label{inv1}
I_2\equiv\sqg(R^\mn P_{,\m}P_{,\n}-\frac{1}{6}RP_{,\m}P^{,\m}
+\frac{3}{4}(\na^2P)^2-\half P^{,\mn}P_{,\mn})\com\nn
I_{1a}\equiv\sqg(R^\mn-\half g^\mn R)P_{,\mn}\com\q
I_{1b}\equiv\sqg\na^2\na^2 P(=\sqg\na^4 P)
                                       \pr
\eea
They transforms under the {\it finite} Weyl transformation as
\bea
\label{inv2}
g'_\mn=\e^{2\al(x)}g_\mn\com\q 
{I_2}'-I_2=2\sqg\na^\m(\al^{,\n}P_{,\m}P_{,\n})\com\nn
{I_{1a}}'-I_{1a}=\sqg\na^\n
\{ 2(P_{,\mn}\al^{,\m}-\na^2P\cdot\na_\n\al)
  -(2\al^{,\m}\al_{,\n}P_{,\m}+\al_{,\m}\al^{,\m}P_{,\n}) \}\com\nn
{I_{1b}}'-I_{1b}=
2\sqg\{ \na^2(\al^{,\m}P_{,\m})-\na_\m(\al^{,\m}\na^2P) \}
-4\sqg\na^\m(\al_{,\m}\al^{,\n}P_{,\n})  
                                       \pr
\eea
$I_2,I_{1a},I_{1b}$ are invariant up to the total derivative
terms which should be compared with the exactly conformal invariants
$I_0$ of (\ref{CI.23}) and $I_4$ of (\ref{s2.12b}). 
This result implies that, even in the presence of the dilaton, 
the structure of the conformal
anomaly given in Sec.2 can be valid if we generalize the
conformal invariants up to the total derivative terms. 
In eq.(\ref{s7}), new "trivial" terms $J_3,J_{2b}$ also
appear.
\bea
\label{tri1}
J_{2b}\equiv\sqg(-R^\mn P_{,\m}P_{,\n}+P_{,\mn}P^{,\mn}-(\na^2P)^2)\com\nn
K_{2b}\equiv\sqg(P^{,\mn}P_{,\mn}+\half (\na^2P)^2)\com\q
g_\mn\frac{\del}{\del g_\mn(x)}\intfx K_{2b}=+2 J_{2b}\com\nn
J_3\equiv\sqg (\na^2P\cdot P_{,\m}P^{,\m}+2P^{,\m}P^{,\n}P_{,\mn})\com\nn
K_3\equiv\sqg P^{,\mn}P_{,\m}P_{,\n}\com\q
g_\mn\frac{\del}{\del g_\mn(x)}\intfx K_3= \half J_3\pr
\eea

\section{Conclusion}
We have presented a formal proof of the conformal
invariance of the conformal anomaly for the non-gauge
case. An argument is made about its validity for
the gauge case, based on the fact that the variation
of the gauge fixed action, under the conformal transformation, 
is a "BRST-trivial" term. We have examined the dilaton's effect
on the above structure by
explicitly calculating the conformal anomaly of the
free photon on the dilaton-gravity background. In this
case, the conformal anomaly is conformally invariant
up to total derivative terms.

\vs 1
\begin{flushleft}
{\bf Acknowledgement}
\end{flushleft}
The author thanks S.Odintsov for 
bringing the problem of the photon-dilaton-gravity anomaly 
to his attention.
He also 
thanks R.Endo and N.Ikeda for discussion and suggestion.
\vs 1


\end{document}